\documentclass[12pt]{article}
\usepackage[left=1in,right=1in,top=1in,bottom=1in,]{geometry}
\usepackage{setspace}
\usepackage{graphicx}
\usepackage{xcolor}
\usepackage{subcaption}
\usepackage{float}

\usepackage{amsmath}
\usepackage{amsthm}
\usepackage{amssymb}
\usepackage{rdd_symbols}
\usepackage{amsfonts}
\usepackage{dsfont}
\usepackage{bm}
\newtheorem{assumption}{Assumption}

\usepackage{tikz-qtree}
\usetikzlibrary{shapes.multipart}
\usetikzlibrary{decorations.pathreplacing}

\usepackage{multirow}
\usepackage{booktabs}
\usepackage{paralist}
\usepackage{enumitem}
\usepackage{outlines}

\usepackage{authblk}

\usepackage[hidelinks]{hyperref} %
\hypersetup{colorlinks, citecolor=, linkcolor=, urlcolor=blue}

\usepackage{natbib}
\newcommand\blfootnote[1]{%
	\begingroup
	\renewcommand\thefootnote{}\footnote{#1}%
	\addtocounter{footnote}{-1}%
	\endgroup
}

\usepackage{authblk}

\title{Discovering Heterogeneous Treatment Effects \\ 
	in Regression Discontinuity Designs}
\author[1,2]{\'{A}goston Reguly}
\affil[1]{Corvinus University of Budapest}
\affil[2]{Georgia Institute of Technology}

\begin{document}
	
	\maketitle
	\blfootnote{\noindent I am grateful for all the guidance and the thorough comments by my former advisor R\'obert Lieli, without him this paper would not exist in this form. I thank two of my Ph.D. examiners G\'abor B\'ek\'es and Michael Knaus for their helpful comments and suggestions. I also thank participants of EEA-ESEM 2021 and IAAE 2021 conferences, Lajos Szab\'o, Arieda Mu\c{c}o, L\'aszl\'o M\'aty\'as, Bal\'azs Vonn\'ak, J\'anos Div\'enyi, Andrea Weber, Sergey Lychagin, T\'imea Moln\'ar and the participants at the Brownbag Seminar Series at CEU for helpful comments and suggestions. The usual disclaimer applies.
		\par 
		\smallskip
		\noindent
		\textit{Email address}: \href{mailto:agoston.reguly@uni-corvinus.hu}{agoston.reguly@uni-corvinus.hu}
		\par 
		\noindent
		\textit{Codes} are available at \url{https://github.com/regulyagoston/RD_tree}
	}
	
	\vspace{-2em}
	\begin{abstract}
		\noindent The paper proposes a causal supervised machine learning algorithm to uncover treatment effect heterogeneity in sharp and fuzzy regression discontinuity (RD) designs. We develop a criterion for building an honest ``regression discontinuity tree'', where each leaf contains the RD estimate of a treatment conditional on the values of some pre-treatment covariates. It is \emph{a priori} unknown which covariates are relevant for capturing treatment effect heterogeneity, and it is the task of the algorithm to discover them, without invalidating inference, while employing a nonparametric estimator with expected MSE optimal bandwidth. We study the performance of the method through Monte Carlo simulations and apply it to uncover various sources of heterogeneity in the impact of attending a better secondary school in Romania.
	\end{abstract}
	
	\vskip 0.5cm
	\begin{description}
		\item {\bf JEL}:  C13, C21, I21
		\item {\bf Keywords}: Regression discontinuity design, regression tree, random forest, heterogeneous treatment effect, CATE, nonparametric estimation.
	\end{description}
	\newpage
	\onehalfspacing
	\vskip 1cm
	
	\section{Introduction}
	In regression discontinuity (RD) designs, one identifies the \emph{average} treatment effect from a jump in the regression function caused by the change in treatment assignment (or the probability of treatment assignment) as a running variable crosses a given threshold. Identification is based on comparing outcomes on the two sides of the cutoff, assuming that all other factors affecting the outcome change continuously with the running variable, which is not manipulable (see, e.g., \citealp{hahn2001identification}, \cite{imbens2008regression}, 
	\cite{lee2010regression},  \citealp{calonico2014robust}, \citealp{cattaneo2022regression}). 
	\par
	This paper contributes to the literature by proposing a machine learning algorithm to discover heterogeneity in the average treatment effect (ATE\footnote{To be more specific it is a conditional average treatment effect, where conditioning refers to be around the cutoff value.}) estimated in an RD setup. The subpopulations that the algorithm searches over are defined by the values of a set of additional pre-treatment covariates.
	Analysis of treatment effect heterogeneity is important for at least two reasons. Firstly, researchers and policymakers gain a more detailed understanding of the treatment by learning the extent to which the treatment works differently in different groups. 
	Indeed, the overall average effect may not be very informative if there is substantial heterogeneity. For example, the treatment may have no impact in one group while having a significant impact in another, or there may even be groups where the average treatment effect has opposite signs.
	Secondly, uncovering treatment effect heterogeneity -- with strong external validity -- can lead to a more efficient allocation of resources. If the budget for implementing a treatment is limited, decision makers can design future policies to focus on treating those groups with the largest expected treatment effects.
	
	Of course, heterogeneity analysis is routinely undertaken in applied work, typically by repeating the main RD estimation within different groups defined by the researcher. Nevertheless, ad-hoc (or even pre-specified) selection of sub-samples has disadvantages: i) when there are many candidate groups defined by pre-treatment covariates, searching across these groups presents a multiple testing problem and without correction, it leads to invalid inference. ii) The relevant groups may have a complicated non-linear relationship with the treatment effect, and discovering the non-linear pattern is cumbersome or impossible ``by hand.'' 
	For example, searching along the interactions of the pre-treatment covariates is usually infeasible, and the researcher only checks a few interactions motivated by theoretical considerations.\footnote{
		\cite{hsu2019testing} carried out a small survey of top publications in economics in 2005 that use the RD design.
		They found that 15 out of 17 papers checked for heterogeneity, and only 2 addressed the issue with interaction terms. The rest use subsample techniques without correcting for multiple testing.} 
	
	By contrast, the method proposed in this paper allows for the discovery of treatment effect heterogeneity systematically based on pre-treatment covariates while offering a solution to the aforementioned challenges. At present, we know of no other paper that accomplishes these goals specifically in an RD setup. The closest paper with an RD focus is perhaps \cite{hsu2019testing}, which develops \textit{tests} for possible heterogeneity in the treatment effect based on the null hypothesis that the conditional average treatment effect (CATE) function is equal to a constant (the overall average treatment effect). Their proposed tests reveal whether there are groups defined in terms of observed characteristics for which the ATE deviates from the overall average, but they leave the discovery and the estimation of the conditional average treatment effect function as an open question. 
	Another recent working paper by \cite{calonico2025heterogeneous} discusses the semi-parametric estimation of the CATE function in RD settings. \cite{calonico2025heterogeneous} develop a framework where the functional form and the source of the treatment heterogeneity are known through observable covariates, while the running variable is estimated nonparametrically. Their contribution establishes existence and probability limits for the CATE function while employing a local polynomial estimator.
	We contribute to the literature by proposing a data-driven machine learning method, which \emph{discovers} groups with different treatment effects, using many candidate pre-treatment variables, without invalidating inference. The method provides discovery in the sense that the researcher does not need to specify the sources of heterogeneity (the relevant variables) in a pre-analysis plan, but can use many potentially relevant pre-treatment variables. The algorithm's task is to find the relevant variables and the functional form from the many possible combinations. The end result gives groups with differences in the treatment effects. The implementation of the algorithm assumes that the standard RD identification conditions hold in the potentially relevant subpopulations; e.g., one cannot consider groups in which the running variable is always above or below the cutoff.
	\par
	The paper also builds on and extends the literature of discovering heterogeneous treatment effects with machine learning methods. There is already an extensive literature using causal supervised machine learning (ML) techniques for this purpose (see, e.g., \citealt{imai2013estimating}, \citealt{athey2016recursive}, \citealt{wager2018estimation}, \citealt{athey2019generalized} \citealt{gnecco2020CTIV}, \citealt{friedberg2020local} or \citealt{alex260169}, \citealt{chernozhukov2024applied}).\footnote{There is another, distinct, strand of the broader causal inference literature where ML techniques are used for estimating high-dimensional nuisance parameters, while the parameter of interest is still the average treatment effect or a reduced dimensional version of CATE. See e.g., \cite{chernozhukov2018}, \cite{chernozhukov2019cate}, \cite{chernozhukov2024applied} or \cite{lieli2020cate}. \cite{Lieli2022} provides a nice overview from the topic.}
	All of these works are concerned with i) randomized experiments and/or ii) observational studies with the unconfoundedness assumption, or iii) using instruments to estimate the local average treatment effect (LATE). E.g., \cite{imai2013estimating} uses lasso with two sparsity constraints to identify heterogeneous treatment effects. The idea is to formulate heterogeneity as a variable selection problem in randomized experiments or observational studies with the unconfoundedness assumption. \cite{athey2016recursive} also focus on randomized experiments or observational studies with unconfoundedness, but use what they call \textit{honest} regression trees to find heterogeneity in the treatment effect. The honest approach means that independent samples are used to grow the tree and estimate the average treatment effect in the resulting leaves. This ensures that traditional confidence intervals constructed for the estimates have the proper coverage rate. 
	\cite{gnecco2020CTIV} follow the \cite{athey2016recursive} approach and extend it with instrumental variable setting to estimate conditional local average treatment effects (CLATE). Finally, the rest of the aforementioned papers and references
	therein go beyond regression trees and use random forests or other machine learning methods to estimate conditional treatment effects in settings i), ii), and iii).\footnote{
		\cite{wager2018estimation}, introduce causal (random) forests and shows that using honest trees to construct the forest, yields asymptotic normality for the conditional treatment effect estimator. They implement their theoretical results for causal forests in randomized experiments or observational studies with unconfoundedness. \cite{friedberg2020local} use `generalized random forests' as an adaptive weighting function to express heterogeneity. 
		\cite{friedberg2020local} improve the asymptotic rates of convergence for generalized random forests with smooth signals by using local linear regressions, where the weights are given by the forests. Their method applies to randomized experiments and shows an application with an observational study, with the unconfoundedness assumption. \cite{knaus2021double} synthesize different methods using double machine learning, focusing on program evaluation under the unconfoundedness assumption. He also proposes a normalized DR-learner to estimate individual average treatment effects. \cite{alex260169} provide a great overview of the Empirical Monte Carlo Study performances of the different machine learning methods available and used in practice.}
	As the aforementioned methods are already developed, one may argue that the CATE function in RD can be estimated by using these causal supervised machine learning methods. However, these methods require some restrictive and unnecessary assumptions when identifying the CATE parameter in the RD context. For example, in observational studies with unconfoundedness, it is impossible to construct a treatment-control contrast without overlap. The lack of overlap causes the propensity score to take either the value of 1 or 0. This is indeed a problem, as in many causal supervised machine learning methods, this is typically excluded by assumption. For example, with propensity score weighting, one needs to assume that the propensity score values are away from the boundaries to make the transformation feasible. Another caveat is that these methods assume that the unobservable factors are the same for treated and control units for every value of the running variable; thus, the conditional expectation functions (CEFs) for both the treated and control units are independent from these unobservables. In contrast, RD only assumes continuity for CEFs around the threshold, and CEFs can be anything away from the cutoff. Another approach is to use instrumental variables, which relaxes the assumption of unobservable factors being the same for both treated and control groups by using instruments. With the introduction of instrument(s), the core assumption is the exclusion restriction; thus, instrumental variables enter only in the selection equation, but not the outcome equation, and are uncorrelated with the unobservables. Usually, it is hard to find such variables in the context of RD. Furthermore, when the instrument is binary, the ATE or CATE can be identified without further assumptions. However, if this is not the case, one must use the ``identification at infinity'' assumption. In contrast, with classical RD design, the researcher can avoid taking such strong assumption(s) while relying on the observed running variable and the continuity assumptions.\footnote{For more detailed discussion on how to estimate ATE with different types of models, see \cite{lee2010regression} Section 3.5.}
	\par
	\indent
	Finally, let us mention a closely related paper by \cite{athey2019generalized}. They work out a general framework for estimating heterogeneous treatment effects called `generalized random forests'  based on local moment conditions. In their paper, they work out the local moment conditions for nonparametric quantile regression, conditional average partial effect estimation, and heterogeneous treatment effect estimation via instrumental variables. However, they do not account for regression discontinuity designs with bandwidth selection. From their perspective, this paper extends their methodology to RD designs with bandwidth selection and derives the properties of a nonparametric RD estimator for a tree and extends to forest.\footnote{For completeness, let us mention a working paper by \cite{ryan2019momentforest}. They use moment-based models when constructing trees, calling it the `moment forest'. They use regression discontinuity design as an application for their method; however, it is in a preliminary state: they make some strong assumptions on the functional form of conditional expectation functions when estimating the CATE function, which we do not require in this paper and use nonparametric method instead.}
	\par
	We contribute to the causal machine learning literature by introducing a specialized machine learning method to search for and estimate conditional average treatment effects in an RD setup. According to our knowledge, there is no paper on combining the estimation of trees or forests to capture heterogeneity with appropriate bandwidth selection for nonparametric RD estimates. Following \cite{athey2016recursive}, we build an honest ``regression discontinuity tree'', where each tree leaf contains a \textit{nonparametric} RD regression. We select the tree(s) jointly with a unified bandwidth that minimizes expected mean squared error. Here, the expectation for the MSE refers to an independently drawn sample from the same population, similarly to \cite{athey2016recursive} and \cite{wager2018estimation}. We derive the nonparametric RD criterion to jointly build a tree and select bandwidth. This criterion contains four parts: expected bias square of the CEF, expected variance of the CEF, expected squared treatment effect, and expected variance of the treatment effect. The bias-variance trade-off for selecting the bandwidth is familiar in the nonparametric RD literature. The expected squared treatment effect and the expected variance will govern the discovery of the differences in the treatment effects. We propose a feasible estimator for each component and analyze its properties.
	Furthermore, we modify the tree-building algorithm in multiple ways to accommodate RD estimation and the new criterion. From a strictly technical standpoint, these are the paper's main contributions.
	\par
	We present Monte Carlo simulations to demonstrate that the algorithm successfully discovers and estimates heterogeneity in various settings --- at least with suitably large samples. In addition, we use the well-known and investigated dataset of \cite{popeleches2013} on the Romanian school system. \cite{popeleches2013} study the average treatment effect on Baccalaureate examination outcomes of going to a better school, and undertake additional ad-hoc heterogeneity analysis. \cite{hsu2019testing} use their proposed test and shows some evidence on the heterogeneity in the treatment effect without identifying the sources of the heterogeneity. We show that using the algorithm, we can refine their results, discovering important treatment heterogeneity along with the level of school average transition scores and the number of schools in town. The algorithm reveals groups with different treatment effects that missed \cite{popeleches2013}. Furthermore, with a more extensive survey dataset with many socio-economic variables (but with fewer observations), we find that the estimated intention-to-treat effect varies among other covariates, including having phone access, school average transition score, negative interaction with peers, and the age of the head of household.
	\vspace{.5em}
	
	\indent
	The paper is organized as follows. Section \ref{sec:theory} introduces the concept of a sharp RD, a regression tree, defines the conditional average treatment effect for the regression discontinuity tree, and discusses the interpretation of CATE. This section briefly overviews extensions to random forest and fuzzy designs. Section \ref{sec:criterion} develops the honest criterion for RD trees, which governs the discovery of the partitions. It also overviews the specifics of employing a nonparametric estimator: unified bandwidth selection, bias, and variance trade-offs. We also discuss some practical guidance on cross-validation, details of honesty in the RD setup, and the order of polynomial selection. Section \ref{sec:simul} shows the Monte Carlo simulation results with a sharp regression discontinuity design for nonlinear running variable cases. Section \ref{sec:emp_app} demonstrates the algorithm's usefulness on datasets collected by \cite{popeleches2013}. Section \ref{sec:Conclusion} concludes.
	%
	\section{Regression Discontinuity Tree}
	\label{sec:theory}
	%
	With sharp regression discontinuity design, researchers are interested in the causal effect of a binary treatment. Let $Y(1)$ denote the potential outcome when a unit gets the treatment and $Y(0)$ denote the potential outcome when no treatment occurs. The observed outcome corresponding to the actual treatment status can be written as
	\begin{equation*}
		Y = Y(D) = \begin{cases}
			Y(0), \qquad \text{if} \quad D=0,\\
			Y(1), \qquad \text{if} \quad D=1 \, .
		\end{cases}
	\end{equation*}
	\noindent
	Treatment assignment in sharp RD is a deterministic function of a scalar variable, called the \textit{running variable}, which is denoted by $X$. This paper considers the standard case, in which the treatment $D$ is determined solely by whether the value of the running variable is above or below a \textit{fixed} and \textit{known} threshold $c$:
	\begin{equation*}
		D = \dc = \mathds{1}_{[c,\infty)}(x)
		\begin{cases}
			1 \,, \quad \text{ if } \quad x \geq c \\
			0\,, \quad \text{ otherwise } \,.
		\end{cases}
	\end{equation*}
	\noindent
	Treatment heterogeneity comes in the form of additional characteristics. Let $Z$ be a set of $K$ random variables referring to the possible sources of heterogeneity. $Z$ are pre-treatment variables; therefore, they must not affect the value of the running variable. Following the machine learning terminology, call these variables \textit{features}. 
	\par
	\indent
	This paper proposes a method to estimate, or in some cases approximate, the conditional average treatment effect function given by
	\begin{equation}
		\label{eq:CATE}
		\tau(z)=\mathbb{E}\left[Y(1)-Y(0)|X=c,Z=z\right] \, .
	\end{equation}
	The proposed regression tree algorithm allows a step-function approximation when this CATE function is continuous in $z$. We also discuss extension to the forest, which can capture continuous effects, but it comes with the cost of additional identification assumptions.
	\subsection{CATE in regression discontinuity tree}
	Regression trees -- sometimes referred to as a partitioning scheme -- allow one to construct a simple, intuitive, and easy-to-interpret step-function approximation to the CATE. A tree $\Pi$ corresponds to a partitioning of the feature space. Partitioning is done by recursive binary splitting: 1) Split the sample into two sub-samples along one feature with a split value. If a unit has a larger value for the selected feature than the split value, then it goes to the first sub-sample; otherwise, it goes to the second sub-sample. 2) If there are multiple candidate splitting value for a given feature check each splitting value and choose which has the lowest value for a pre-determined criterion (e.g. MSE) 3) Iterate through each feature and pick the variable and split value that has the lowest criterion value. 4) If needed, one repeats the split, but now one considers the already split sub-samples for the next split. This way, the feature space $(\mathds{Z})$ is partitioned into mutually exclusive rectangular regions. In the \href{https://github.com/regulyagoston/RD_tree/blob/main/online_appendix.pdf}{online appendix}, Figure A.1 shows illustrative examples, with two features $Z_1$ and $Z_2$. 
	\par 
	The final regions are called `\textit{leaves}' or `\textit{partitions}', denoted by $\ell_j$, where $\ell_j \subseteq \mathds{Z}$ and $\ell_j \cap \ell_{j'} = \emptyset \,, \forall j\neq j'$. A regression tree, $\Pi$ has $\nP$ leaves, $j=1,\dots,\nP$, whose union gives back the complete feature space $\mathds{Z}$. 
	\begin{equation*}
		\Pi = \{\ell_1,\dots,\ell_j,\dots,\ell_{\#(\Pi)}\}, \quad \text{with} \qquad \bigcup_{j=1}^{\#\Pi}\ell_j=\mathds{Z}
	\end{equation*}
	\noindent
	Recursive splitting provides rectangular regions for the different treatment effects, but never a continuous function. In the case of a continuous CATE, a simple tree offers only a step-function approximation. However, the tree structure ensures an intuitive decision-based interpretation of the treatment effects.
	Until Section \ref{sec:criterion}, let us assume that the tree $\Pi$ is given, however, for later sections, when there are multiple candidate trees, let us use the expression $\ell_j(\Pi)$ instead of simply using $\ell_j$. The average treatment effect for leaf $\ell_j(\Pi)$ is defined as
	\begin{equation}
		\label{eq:CATE_ell_j}
		\tau_j = \E\left[Y(1)-Y(0)|X=c,Z \in \ell_j(\Pi)\right] \, .
	\end{equation}
	\noindent
	To state the regression discontinuity tree approximation to the whole CATE function, let us introduce the indicator function for leaf $\ell_j(\Pi)$,
	\begin{equation*}
		\dl =
		\begin{cases}
			1 \,, \quad \text{ if } \quad z \in \ell_j(\Pi) \\
			0\, \quad \text{ otherwise } \,.
		\end{cases}
	\end{equation*}
	\noindent
	The approximated conditional average treatment effect function provided by the regression discontinuity tree is given by
	\begin{equation}
		\label{eq:CATE_RD}
		\tau(z;\Pi) = \sum_{j=1}^{\nP} \tau_j \dl \, .
	\end{equation}
	This CATE function -- which incorporates the tree structure -- links the treatment effects for each leaf. As the leaves represent rectangular partitions, this function is a step-function approximation to the continuous CATE function. By the law of iterated expectation, this approximation has the property of $\E\left[\tau(Z)\bigm\vert \mathds{1}_{\{c\}}(X) ,\mathds{1}_{\ell_1}(Z),\dots,\mathds{1}_{\ell_\nP}(Z)\right] = \tau(Z;\Pi)$. This means that at the threshold value $(X=c)$ with the given tree structure, the expected value of the continuous CATE function over the leaves is equal to the step-approximated CATE.
	\subsection{Identification of CATE in the sharp RD}
	To identify the conditional average treatment effect function for trees in sharp RD, the following assumptions are needed:
	\vspace{1em}
	\begin{assumption}[Identification assumptions]~ 
		\label{ass:CondCov}
		\begin{enumerate}
			\item[a)] $\mathbb{E}\left[Y(1)|X=x,Z \in \ell_j(\Pi)\right]$ and $\mathbb{E}\left[Y(0)|X=x,Z \in \ell_j(\Pi)\right]$, exists and continuous at $x=c$ for all leaves in the tree.
			\item[b)] Let $f_j(x)$ denote the conditional density of $x$ in leaf $j$. In each leaf $j$, $c$ is an interior point of the support of $f_j(x)$.
		\end{enumerate}
	\end{assumption}
	%
	\noindent
	Assumption~\ref{ass:CondCov}a) states that the expected value of the potential outcomes conditional on the running variable in each leaf exists and continuous. It is required to identify the average treatment effects for all leaves. This assumption is similar to the classical RD assumption (see e.g., \citealt{imbens2008regression}), but somewhat stronger, due to extension to the tree.\footnote{This is less restrictive if one assumes continuity in $Z=z$ as in e.g., \cite{hsu2019testing} or as we will need for discontinuity forest.} Assumption \ref{ass:CondCov}b) ensures that the density for the running variable is well behaved: it has a positive probability below or above the threshold value within each leaf. This excludes cases when there are no values of the running variable on both sides of the threshold in a given leaf. Finally, in the RD literature, it is common to require the continuity of the conditional distribution functions -- in this case, it extends to $f_j(x)$ to be continuous in $x$\footnote{Bayes' rule along with assumption a) gives this result.} -- which is an implication of \textit{``no precise control over the running variable''} (see e.g., \citealt{lee2010regression}). The algorithm does not need this assumption if local randomization around the threshold holds.
	\par
	\noindent
	Under these assumptions, the (step-function approximated) conditional average treatment effect given by a regression discontinuity tree is identified as
	\begin{equation}
		\label{eq:id_cate}
		\begin{aligned}
			\tau(z;\Pi) 
			&= \sum_{j=1}^{\nP} \tau_j \dl \\
			&= \sum_{j=1}^{\nP} \left\{
			\mathbb{E}\left[Y(1)|X=c,Z \in \ell_j(\Pi) \right]     
			-\mathbb{E}\left[Y(0)|X=c,Z \in \ell_j(\Pi) \right] 
			\right\} \,\, \dl \\
			&= \sum_{j=1}^{\nP} \left\{
			\lim\limits_{x \downarrow c} \mathbb{E}\left[Y(1)|X=x,Z \in \ell_j(\Pi) \right]
			- \lim\limits_{x \uparrow c}\mathbb{E}\left[Y(1)|X=x,Z \in \ell_j(\Pi) \right] 
			\right\} \,\, \dl \\
			&= \mu_+(c,z;\Pi) - \mu_-(c,z;\Pi)\\
		\end{aligned}
	\end{equation}
	where
	\begin{equation}
		\label{eq:con_exp}
		\begin{aligned}
			\mu_+(x,z;\Pi) &= \sum_{j=1}^{\nP} \mathbb{E}\left[Y(1)|X=x,Z \in \ell_j(\Pi) \right] \,\, \dl \\
			\mu_-(x,z;\Pi) &= \sum_{j=1}^{\nP} \mathbb{E}\left[Y(0)|X=x,Z \in \ell_j(\Pi) \right] \,\, \dl
		\end{aligned}
	\end{equation}
	\noindent
	refers to the conditional expectation function for ($\mu_+$)~above the threshold (treated) and ($\mu_-$)~below the threshold (untreated) units. That is, each $\tau_j$ is identified within its leaf in the usual way.
	\subsection{Interpretation of the estimand conditional on (un)observables}
	The conditional average treatment effect estimand in regression discontinuity designs is not as straightforward as in experimental designs or observational studies with the unconfoundedness assumption. To interpret the estimand, let us formalize the individual treatment effect following \cite{lee2010regression},
	\begin{equation*}
		Y(1) = Y(0) + \tau(Z,U)
	\end{equation*}
	where $Z$ are the known observed covariates and $U$ is unobserved heterogeneity in the individual treatment effect. Let us consider the simple sharp RD setup, where $\tau(Z,U)$ does not depend directly on the running variable $X$. Note that $X$, $Z$, and $U$ can be correlated; thus, individuals with characteristics of $Z$ and $U$ can have typical $X$ values, but $X$ does not directly influence the magnitude of the treatment effect.
	\par
	\noindent
	Naturally, individual treatment effects can not be observed as one can not simultaneously assign the same unit to be treated and non-treated. Instead, one can identify a type of conditional average treatment effect, where $Z$ and $X$ are fixed and $U$ is averaged out:
	\begin{equation*}
		\tau(z) = \E\left[ \tau(Z,U) \mid X = c , Z = z \right] = \E\left[Y(1) - Y(0) \mid X = c , Z = z \right]  \, .
	\end{equation*}
	\par
	\noindent
	We consider the case when $Z$ and $U$ are discrete for interpretation purposes, but a similar argument applies to the continuous case. First, focus on the general case when no tree structure is used. CATE is identified through
	\begin{equation*}
		\tau(z) = \lim_{x \downarrow c} \E \left[ Y \mid X = x, Z = z \right] 
		- \lim_{x \uparrow c} \E \left[ Y \mid X = x, Z = z \right]  \, .
	\end{equation*}
	Under the identifying Assumption \ref{ass:CondCov}a) and \ref{ass:CondCov}b), the CATE function is equal to,
	\begin{equation*}
		\begin{aligned}
			\E\left[Y(1) - Y(0) \mid X = c , Z = z \right] 
			&= \E\left[ \tau(Z,U) \mid X = c , Z = z \right] \\
			&= \sum_u \tau(z,u)\Prb\left[U=u \mid X=c,Z=z\right]\\
			&= \sum_u \tau(z,u)\frac{f(c \mid U=u,Z=z)}{f(c\mid Z=z)}\Prb\left[U=u \mid Z=z\right]
		\end{aligned}
	\end{equation*}
	where $\Prb\left[\cdot \mid \cdot\right]$ denotes conditional probability and $f(\cdot \mid \cdot)$ denotes conditional density function.\footnote{This formula is the exact analog of Equation (5) in \cite{lee2010regression}.}
	Thus, the CATE function is a particular kind of average treatment effect
	across individuals with covariate values $Z=z$. If the term $f (c | U = u, Z = z)/f (c \mid Z = z)$ were equal to 1, it would be the treatment effect for individuals with observed $Z=z$ averaged over the unobserved $U=u$ values. This is the case if the unobserved heterogeneity $U$ is independent of the running variable $X$ conditional on the covariates $Z$. More generally, the presence of the ratio $f (c | U = u, Z = z)/f (c \mid Z = z)$ implies the regression discontinuity estimand is instead a weighted average treatment effect.
	Within the subgroup $Z=z$, the weight is larger for individuals whose $X$ value is ex-ante more likely to be close to the threshold $c$ based on their unobserved characteristics.
	The weights may be relatively similar across individuals, in which case the individual treatment effects would be closer to the CATE, but if the weights are highly varied and also related to the magnitude of the treatment effect, then the individualized treatment effects would be very different from the CATE. However, the weights across individuals are ultimately unknown, since we do not observe $U$. Thus, it is not possible to know how close the individualized treatment effects are to the CATE, and it remains the case that the treatment effect estimated using an RD design is averaged over a larger population than one would have anticipated from a purely ``cutoff'' interpretation. 
	\par
	Finally, let us discuss the impact of using a regression tree representation in the interpretation of the CATE function. Following Equation (\ref{eq:CATE_ell_j}) from the paper, the leaf-by-leaf treatment effect can be similarly decomposed to
	\begin{equation*}
		\begin{aligned}
			\tau_j 
			&= \E\left[Y(1)-Y(0) \mid X = c, Z \in \ell_j \right] \\
			&= \E\left[\tau(Z,U) \mid X = c, Z \in \ell_j \right] \\
			&= \sum_{z,u} \tau(z,u) \Prb\left[ Z=z,U=z \mid X=c, Z\in \ell_j \right] \\
			&= \sum_{z \in \ell_j,u} \tau(z,u) \frac{f(c \mid Z=z, U=u, Z \in \ell_j )}{f(c \mid Z \in \ell_j)} \Prb\left[ Z=z,U=u \mid Z\in \ell_j \right] \, .
		\end{aligned}
	\end{equation*}
	The interpretation remains similar, but with a tree structure, one needs to average over not only the unobserved characteristics $(U=u)$, but over the observed characteristics within each leaf $j$ as well.
	\par
	Note that
	if there is no unobserved heterogeneity in the treatment effect $(\tau(Z,U)=\tau(Z))$, then in the continuous case, one can estimate the individualized treatment effects. With a tree structure, weights are still present as the conditional densities are not necessarily the same within leaf $j$ for each value of $z$.
	Also, if the tree specification is correct in the sense that $\E\left[\tau(Z,U) \mid X = c, Z=z \right]=\E\left[\tau(Z,U) \mid X = c, Z \in \ell_j \right]$, then the interpretation is the same as if $\tau(Z,U)$ would be continuous in $Z$.
	Finally, if the tree is correctly specified and there is no unobserved heterogeneity in the treatment effect, then the CATE via tree structure is the same as the individualized treatment effect.
	%
	\subsection{Extension to discontinuity forest}
	\label{sec:rf-identification}
	As shown before, the discontinuity tree gives a step-approximated CATE if it is continuous in pre-treatment variable(s) $Z$. Building on the results of \cite{wager2018estimation}, the causal forest extension of the discontinuity tree is a natural way to estimate continuous CATE. We require same assumptions as \cite{wager2018estimation}: CEFs are both Lipschitz continuous and for asymptotic sampling distribution that the sub-sample size $s$ scales as $s \asymp N^\xi$ for some $\xi_{min} < \xi < 1$, where $N$ is the sample size. Instead of the overlap assumption we need to extend Assumptions \ref{ass:CondCov}a) and \ref{ass:CondCov}b). Specifically it is required $\mathbb{E}\left[Y(d)|X=x,Z=z\right]$ exists and continuous at $x=c$, for all $Z=z$, and $d=\{0,1\}$. Furthermore, let $f_{Z=z}(x)$ denote the conditional density of $x$ for all $Z=z$. We need that for all $z$, that $c$ is an interior point of the support of $f_{Z=z}(x)$. In practice, one need to be careful, if these assumptions are too restrictive or not with many covariates.

	\subsection{Identification with fuzzy design}
	\label{sec:fuzzy_id}
	The method can be extended to fuzzy designs as well, where the probability of treatment does not need to change from 0 to 1 at the threshold and can allow for a smaller jump in the probability of assignment.
	\par
	\indent
	Let us use a distinct variable $T$ to get the treatment in the case of fuzzy design. As the probability does not change from 0 to 1 at the threshold, there are different types of participants, depending on whether they are subject to the treatment or not. Compliers are units that get the treatment if they are above the threshold but do not get the treatment if they are below: $T(1)-T(0)=1$. Always takers get the treatment regardless of whether they are below or above the threshold, while never takers never take the treatment regardless of the threshold value. For both behaviors, the following applies $T(1)-T(0)=0$. As in classical fuzzy RD, we eliminate by assumption defiers, who do not take the treatment if above the threshold and take the treatment if below the threshold.
	\par
	\indent
	Fuzzy RD identifies treatment effects for compliers, thus extending the algorithm to fuzzy designs, resulting in conditional local average treatment effects (CLATE). To identify CLATE, the following assumptions are needed:
	\vspace{1em}
	\begin{assumption}[Identification assumptions for fuzzy design]~ 
		\label{ass:fRD}
		\begin{enumerate}    
			\item[a)] $\lim_{ x \downarrow c}\mathbb{P}\left[T=1|X=x\right] \geq \lim_{ x \uparrow c}\mathbb{P}\left[T=1|X=x\right]$
			\item[b)] $\E\left[Y(d) \, | \, T(1)-T(0)=d',X=x,Z \in \ell_j(\Pi)\right]$ exists and continuous at $x=c$ for all pairs of $d,d' \in \{0,1\}$ and for all leaves $j$ in the tree.
			\item[c)] $\mathbb{P}\left[T(1)-T(0) = d \, | \, X = x, Z \in \ell_j(\Pi) \right]$ exists and continuous at $x=c$ for $d \in \{0,1\}$, $\forall j$ and for all leaves $j$ in the tree.
			\item[d)] Let, $f_j$ denotes the conditional density of $x$ in leaf $j$. In each leaf $j$, $c$ must be an interior point of the support of $f_j(x)$.
		\end{enumerate}
	\end{assumption}
	\vspace{1em}
	\noindent
	Identification assumptions are similar to classical fuzzy RD but must be valid within each leaf. Assumption \ref{ass:fRD}a) rules out defiers as it requires a non-negative discontinuity in the probability of taking the treatment around the threshold. This is not only an assumption, but a built-in restriction for the algorithm. If this condition's sample analogue is not satisfied, it is not considered a valid split. Assumptions \ref{ass:fRD}b) and \ref{ass:fRD}c) ensure the existence and continuity of the expected potential outcomes at the threshold value for always-takers, compliers and never-takers with respect to the running variable within each leaf, while assumption \ref{ass:fRD}d) ensures that the conditional density of $x$ for each leaf is well behaving, similarly to sharp RD.
	\noindent
	Under these assumptions, the CLATE for RD tree is identified as
	\begin{equation*}
		\begin{aligned}
			\tau_{FRD}(z;\Pi) = 
			\frac{\lim_{ x \downarrow c} \mu_+^y(x,z;\Pi) - \lim_{ x \uparrow c} \mu_-^y(x,z;\Pi)}
			{\lim_{ x \downarrow c} \mu_+^t(x,z;\Pi) - \lim_{ x \uparrow c} \mu_-^t(x,z;\Pi)} 
			= \frac{\mu_+^y(c,z;\Pi) - \mu_-^y(c,z;\Pi)}
			{\mu_+^t(c,z;\Pi) - \mu_-^t(c,z;\Pi)} \,,
		\end{aligned}
	\end{equation*}
	\noindent
	where, $\mu$ is the conditional expectation function similarly defined to Equation (\ref{eq:con_exp}), but now in the superscript we denote the outcome ($y$), or the participation $(t)$.
	
	\section{Estimation}
	\label{sec:criterion}
	First, we introduce different distinct samples necessary to obtain an unbiased estimator of the CATE function when using the regression tree algorithm. Secondly, we analyze the criterion, which compares different candidate trees. 
	Third, we introduce a q-th order local polynomial in $X$ to estimate $\tau_j$ and derive a feasible criterion. (We also derive a parametric estimator in \href{https://github.com/regulyagoston/RD_tree/blob/main/online_appendix.pdf}{online appendix}, under Section D. that allows to analyze a simpler version of the criterion.) After analyzing the criterion, we discuss the fuzzy design, provide refinements for discontinuity trees, and extend to discontinuity forests.
	\subsection{Distinction of samples}
	An inherent problem of using only one sample for finding relevant sub-groups and estimating treatment effects is that it results in incorrect inference if there is no adjustment for multiple testing. (see, e.g., \citealp{anderson2008multiple})
	\par
	\indent
	Although the regression tree algorithm controls for over-fitting in some way, the estimate is biased in finite samples and disappears only slowly as the sample size grows. \cite{athey2016recursive} was the first paper to propose `\textit{honest regression tree}' approach, which eliminates the bias from the estimated conditional average treatment effects in experimental settings or observational studies with the unconfoundedness assumption. By their definition, a regression tree is called `honest' if it does not use the same information for growing the candidate trees as for estimating the parameters of that tree. This requires using two \textit{independent} samples. The `\textit{test sample}' $(\ste)$ is used for evaluating the candidate trees, and the `\textit{estimation sample}' $(\sest)$ for estimating the treatment effects. These samples are also used to derive and analyze the honest criterion for the regression discontinuity tree. 
	Honesty implies that the asymptotic properties of treatment effect estimates within the partitions are the same as if the partition had been exogenously given; thus, biases are eliminated, and one can conduct inference in the usual way. The cost of the honest approach is the loss in precision -- less observation used -- due to sample splitting \citep[p. 7353-7354]{athey2016recursive}.\footnote{With the honest approach, one does not need to place any external restrictions on how the tree is constructed. In the literature, other papers use additional assumptions to get valid inference. An example is \cite{imai2013estimating}, which use the `sparsity' condition: only a few features affect the outcomes.}
	\subsection{Criterion for RD tree}
	A natural -- but infeasible criterion -- for evaluating the regression discontinuity tree would be minimizing the mean squared error of the estimated CATE on the test sample. First, let a partition $(\Pi)$ and bandwidth $(h)$ be exogenously given, and later we relax these assumptions. Consider any sample $\mathcal{S}$, consisting of independent and identically distributed observations $(Y_i,X_i,Z_i)\,;i = 1, \dots, N$. The estimated CATE function, $\tllestih$, is estimated on $\sest$ and evaluated on $\ste$. The infeasible MSE criterion is
	\begin{equation}
		\label{eq:infMSE}
		MSE_\tau(\ste,\sest,\Pi,h) = \frac{1}{N^{te}}\sum_{i\in\ste}\left\{\left[ \ttau - \tllestih \right]^2 - \tau^2(Z_i) \right\}
	\end{equation}
	where $N^{te}$ is the number of observations in the test sample and $Z_i$ refers to features also from $\ste$. Note, this formulation has an extra adjustment term, $\tau^2(Z_i)$ -- a scalar, independent of $\Pi$. Thus, it does not affect the results but facilitates theoretical derivations. Furthermore, let us emphasize that this infeasible criterion utilizes both the estimation and test samples in a way that observations need to be known for both samples.
	\par
	Calculating this criterion for different exogenously given trees and bandwidths would allow one to find the tree and bandwidth whose deviation from the true CATE function is the smallest in the test sample. The problem is $\tau(\cdot)$ is unknown; thus, this criterion is infeasible.
	Instead -- following \cite{athey2016recursive} -- we minimize the \textit{expected} MSE over the test and estimation samples. This formulation has two advantages: i) it gives the best fitting tree and bandwidth for the \textit{expected} test and estimation sample. This is favorable because when the tree is grown and bandwidth is chosen, both samples are locked away from the algorithm; ii) using this formulation, an estimable criterion can be derived for comparing trees and bandwidths. The expected MSE criterion is given by
	\begin{equation}
		\label{eq:EMSE}
		EMSE_\tau(\Pi,h) = \E_{\ste,\sest} \left[MSE_\tau(\ste,\sest,\Pi,h)\right] \, .
	\end{equation}
	This paper advocates trees $(\Pi)$ with unified bandwidth $h$, which gives the smallest $EMSE_\tau$ value from all the candidates.\footnote{An alternative for bandwidths is to use leaf-by-leaf bandwidth $h_j$, that is MSE/CE optimal for each leaf. This option is also available in the implementation; however, our simulation result showed that this method has worse sample properties than the unified version.} 
	The EMSE criterion can be decomposed into four terms, see the derivations in the \href{https://github.com/regulyagoston/RD_tree/blob/main/online_appendix.pdf}{online appendix}, under Section B.1.
	\begin{equation} 
		\label{eq:sEMSE}
		\begin{aligned}
			EMSE_\tau(\Pi,h)  &= \E_{Z_i} \left\{ \E_{\sest} \left[ \mathbb{B}^2(\hat\tau(z;\Pi,h,\sest))\right] \bigm\vert_{z=Z_i} \right\} + \E_{Z_i} \left\{ \E_{\sest} \left[ \V(\hat\tau(z;\Pi,h,\sest)) \right] \bigm\vert_{z=Z_i} \right\} \\
			&- \E_{Z_i}\{\E_{\sest}[\tilde\tau^2(Z_i;\Pi,h,\sest)]\}+\E_{Z_i}\{\V_{\sest}[\tilde\tau(Z_i;\Pi,h,\sest)]\}\,.
		\end{aligned}
	\end{equation}
	The first line represents the squared bias and the variance of the CATE estimator evaluated for each $z=Z_i$ and then takes the expectation over the observed test sample. This part shows the well-known bandwidth selection problem in RDD,\footnote{There is a large literature on how to select the bandwidth for local polynomial estimators in the RD setting, see e.g., \cite{calanico2020optbw}.} however, let us emphasize that our criterion uses the expectations over $z=Z_i$, thus the resulting bandwidth is going to be optimal over the population and not for each leaf.
	\par
	The second line highlights the trade-off between finding new, different treatment effects (the squared term of the CATE) and minimizing the variance of the estimated treatment effects that is familiar in the causal supervised machine learning literature. Note, however that we used $\tilde{\tau}$ instead of $\hat\tau$, to highlight bias-corrected estimator for treatment effect, which we discuss in the next section. The expected value of the squared CATE term prefer larger trees, as the expected squared treatment effects grow as there are more leaves (or groups). On the other hand, an increasing number of splits leads to too large trees, i.e., where the treatment effects are the same in different leaves. This is called over-fitting. The fourth terms explicitly incorporate that finer partitions generate greater variance in leaf estimates with finite samples. Therefore, it prefers smaller trees, where the average variance of the estimated treatment effects is lower.
	Through these channels, these terms dampen the over-fitting caused by the first and third terms. 
	Also, note that the expected variance (fourth) term may select larger trees if leaves (or groups) have the same treatment effect, but have lower expected variances separately. 
	\par
	Finally, let us highlight that each term depends on the tree structure and the bandwidth. $\Pi$ is related to the complexity of the CATE function, and $h$ is related to the curvature of each conditional expectation function at the threshold (for each leaf and for below and above the threshold). This criterion balances between discovering different treatment effects and different functional forms. For example, suppose a small portion of the population has slightly different treatment effects, but the curvature of the CEFs are vastly different at the treshold, then due to the unified bandwidth, both the expected squared bias and expected variance of the CATE estimator increase more than the squared treatment effect. This results in a worse criterion, thus a smaller tree. Similarly, if a smaller bandwidth results in smaller squared bias and slightly larger expected variance, but with larger squared treatment effect differences, then the algorithm prefers this combination of bandwidth and tree, resulting in a larger tree. This interaction of the four terms is new in the literature.
	
	\subsection{Estimator for EMSE criterion}
	In the RD literature, nonparametric estimation via local linear or local polynomial estimation has become standard (see, e.g., \citealp{calonico2014robust}, \citealp{cattaneo2022regression}). We estimate each conditional expectation function -- $\mathbb{E}\left[Y(d)|X=x,Z \in \ell_j(\Pi) \right], d \in \{0,1\}$ -- with a $q$-th order local polynomial estimator, thus within each leaf the conditional expectations might differ. 
	\par
	\noindent
	First let us adjust $X$ by $c$, and let $\bm{X}$ be the $(q+1) \times 1$ vector, 
	\begin{equation*}
		\bm{X} = \left[ 1 , (X-c) , (X-c)^2 , \dots , (X-c)^q \right]' \,.
	\end{equation*}
	We use one bandwidth $h$ and the same kernel $k(\cdot)$. For each leaf we have separate parameter vectors $\djp= \left[ \alpha_j^+ , \beta_1^+ , \beta_2^+ , \dots , \beta_q^+ \right]'$ and $\djm= \left[ \alpha_j^- , \beta_1^- , \beta_2^- , \dots , \beta_q^- \right]'$ which are both $(q+1) \times 1$.\footnote{For RD, the main parameter of interest is $\alpha_j^\pm$. $\beta^\pm$ should also be $\beta_1,j^\pm$, but we neglect $j$ subscript for convenience.} 
	The nonparametric leaf-by-leaf estimator is given by,
	\begin{equation}
		\label{eq:nonparametricOLS}
		\begin{aligned}
			\hat{\bm{\delta}}_{j,+} &= \arg\min_{\djp} \,\, \sum_{ i \in \mathcal{S}} 
			\left\{ \dci \dli \left( Y_i - \x_i  \djp \right)^2 k\left(\frac{X_i-c}{h}\right) \right\} \\
			\hat{\bm{\delta}}_{j,-} &= \arg\min_{\djm} \,\, \sum_{ i \in \mathcal{S}} 
			\left\{  \left[ 1 - \dci \right] \dli \left( Y_i - \x_i  \djm \right)^2 k\left(\frac{X_i-c}{h}\right) \right\} 
			\,\, , \qquad  \forall j \, .
		\end{aligned}
	\end{equation}
	\noindent
	Note that the subscript $i$ refers to observations from sample $\mathcal{S}$, the $j$ index represents leaf $j$ from tree $\Pi$, and the subscripts $+/-$ stand above or below the threshold.
	\par
	The naive local polynomial estimator for conditional average treatment effect would be $\hat{\tau}(z;\Pi,h,\mathcal{S}) = \hat{\mu}_+(c,z;\Pi,h,\mathcal{S}) - \hat{\mu}_-(c,z;\Pi,h,\mathcal{S}) = \sum_{j=1}^{\nP} \dl \left( \hat \alpha_{+,j} - \hat \alpha_{-,j} \right)$. Notice that these quantities are dependent on the bandwidth $h$. As $\hat{\tau}(z;\Pi,h,\mathcal{S})$ is a biased estimator for $\tau(z)$ we follow \cite{calonico2014robust}, who proposes a bias-corrected estimator 
	\begin{equation*}
		\tilde{\tau}(z;\Pi,h,\mathcal{S}) = \hat{\tau}(z;\Pi,h,\mathcal{S}) - \hat{\mathcal{B}}(z;\Pi) h^2 = \sum_{j=1}^{\nP} \left\{ \dl \left( \hat \alpha_{+,j} - \hat \alpha_{-,j} \right) - \hat{\mathcal{B}}_j h^2 \right\} \,,
	\end{equation*}
	\noindent
	where $\hat{\mathcal{B}}_j$ is the estimated nonparametric bias in leaf $j$ using local polynomial estimator with order $q+1$. In the following, we utilize the results of \cite{calonico2014robust} to get estimators for each part of the EMSE criterion defined by Equation (\ref{eq:sEMSE}). Derivations and expressions of each quantities are provided in the \href{https://github.com/regulyagoston/RD_tree/blob/main/online_appendix.pdf}{online appendix} Section B.2.\footnote{For the derivations we have used two further simplifying assumptions: i) the share of observations within each leaf -- the number of observations within the leaf compared to the number of observations in the sample -- are the same for the estimation and test sample. ii) the shares of units below and above the threshold within each leaf are the same for the estimation and test sample. Asymptotically, both assumptions are true.}
	\par
	The first term refers to the expected squared bias, which is given by
	\begin{equation}
		\label{eq:EMSE_exp_bias_z}
		\hat\E_{Z_i} \left\{ \hat\E_{\sest} \left[ \hat{\mathbb{B}}^2(\hat\tau(z;\Pi,h,\sest)) \right]\bigm\vert_{z=Z_i} \right\} = h^{2q+2} \sum_{j=1}^{\#\Pi}\hat\Bc^2_j p_j^{est} \,,
	\end{equation}
	where $p_j^{est}$ is the share of observation from the estimation sample within leaf $j$. Notice this term increases in $h$. Furthermore, if bias in leaf $j$ is specific to the test sample, but not to the estimation sample, $p_j^{est}$ will be small, which helps to avoid overfitting. The second term is the expected variance for $z=Z_i$ over the estimation sample. Our derived estimator is given by
	\begin{equation}
		\label{eq:EMSE_exp_var_z}
		\hat\E_{Z_i} \left\{ \hat\E_{\sest} \left[ \hat\V(\hat\tau(z;\Pi,h,\sest)) \right] \bigm\vert_{z=Z_i} \right\}
		= \frac{1}{N^{est}h} \sum_{j=1}^{\#\Pi}\hat\Vc_j\,,
	\end{equation}
	where $N^{est}$ is the number of observations in the estimation sample, and $\hat\Vc_j$ is a scalar related to the (raw) variance.\footnote{Raw in the sense of no bias-correction is used here. This is a scaler that is composed of the inverse of the weighted variance-covariance matrix of $\bm{X}$ and the error variance within leaf $j$. See more in \href{https://github.com/regulyagoston/RD_tree/blob/main/online_appendix.pdf}{online appendix} Section B.3.} Note that this measure is decreasing in both $h$ and $N^{est}$.
	\par
	The estimator for the expected value of the squared CATE (third term of Equation \ref{eq:sEMSE}) uses only the test sample,
	\begin{equation}
		\label{eq:EMSE_exp_tau_sq}
		\hat\E_{Z_i}\{\E_{\sest}[\tilde\tau^2(Z_i;\Pi,h,\sest)]\} = \frac{1}{N^{te}}\sum_{i\in \ste}\tilde{\tau}^2(Z_i;\Pi,h,\ste) 
	\end{equation}
	and naturally given by the average of the square of the bias-corrected CATE estimator. Note that this term prefers trees with many leaves. It is the sample analog for finding groups with different treatment effects. This term always increases as the number of leaves increases, while the average of the sum of squared treatment effects for two (or more) groups is always greater than the average of the sum of one averaged squared treatment effect. 
	Finally, our proposed estimator for the expected variance (fourth term from Equation \ref{eq:sEMSE}) is
	\begin{equation}
		\label{eq:EMSE_exp_bc_var}
		\E_{Z_i}\{\V_{\sest}[\tilde\tau(Z_i;\Pi,h,\sest)]\} = \sum_{j=1}^{\nP}p^{te}_j \V\left[\tilde\tau_j(h,\Pi)\right]  \,.
	\end{equation}
	This term uses each leaf's bias-corrected variance estimator for $\tilde{\tau}_j(h,\Pi)$. Note that the scaling for the average comes from the test sample, as the estimator refers to the expected value over the test sample; hence, we use $p^{te}_j$.\footnote{An alternative estimator would be using the estimation sample only. However, our goal is to construct an EMSE estimator, which uses only the test sample's observation and only some additional information from the estimation sample, to ensure that during the tree-building phase, the estimation sample is locked away to get valid inference.}
	\par
	\vskip .3em
	\noindent
	\textit{Remarks on estimators: } 
	\begin{enumerate}[topsep=0pt,itemsep=-1ex,partopsep=1ex,parsep=1ex]
		\item [i)] Although the variance of the treatment effects refers to the estimation sample, we can use observations only from the test sample. This is possible, as the estimation and the test samples are independent. Therefore, the asymptotic estimators for these quantities are the same. 
		\item [ii)] To adjust the variance estimator in finite samples for the estimation sample, one only needs to use limited information from the estimation sample, namely the share of observations $(p^{est}_{j})$.
		\item [iii)] Using the leaf shares instead of the number of observations is possible, as the variance of the treatment effect estimators is the same for each observation within the leaf; therefore, one can use summation over the leaves $(j=1,\dots,\nP)$ instead of individual observations.
	\end{enumerate}
	\par
	\bigskip
	\noindent
	Putting together the estimators, one gets the following estimable EMSE criterion for regression discontinuity trees:
	\begin{equation}
		\label{eq:estEMSE}
		\begin{aligned}
			\widehat{EMSE}_\tau(\ste,\sest,\Pi,h) 
			=& h^{2q+2} \sum_{j=1}^{\nP} \hat{\mathcal{B}}_j^2 p^{est}_j + \frac{1}{N^{est}h}\sum_{j=1}^{\nP} \hat\Vc_j \\
			& -\frac{1}{N^{te}}\sum_{i\in \ste}\tilde{\tau}^2(Z_i;\Pi,h,\ste)
			+ \sum_{j=1}^{\nP}p^{te}_j \V\left[\tilde{\tau}_j(h,\Pi)\right] 
		\end{aligned}
	\end{equation}
	\par
	\noindent
	Minimizing this criterion leads to bandwidth with minimizing the \textit{average} squared bias and \textit{average} variances. Also it leads to trees, where there is strong evidence for heterogeneity in the treatment effects for different groups, and penalizes a partition that creates variance in leaf estimates. Furthermore, this criterion encourages partitions where the variance of a treatment effect estimator is lower, even if the leaves have the same average treatment effect, thus finding features that affect the mean outcome but not the treatment effects themselves.
	\par
	\indent
	Next, let us compare the estimator for the EMSE criterion and the initial infeasible MSE criterion. As the infeasible MSE criterion uses the estimation sample to get an estimator for the CATE function and then evaluates it on the test sample, the estimator for EMSE criterion uses the observations from the test sample and only scales it with the number of observations $(N^{est})$ and share of units for each leaf $(p^{est}_{j})$ from the estimation sample. This means that only limited information is needed from the estimation sample to calculate the EMSE criterion, not individual observations. This property allows honesty: observation values from the estimation sample are locked away for the algorithm when searching for an optimal tree.
	\par
	Let us briefly discuss the connection of this criterion to well known MSE-optimal bandwidth selection from the nonparametric literature, which solves for $arg\min_{h}\left(h^{2q+2}\mathcal{B}^2+\frac{1}{Nh}\mathcal{V}\right)$.\footnote{See e.g., \cite{imbensKalyanaraman2012} or \cite{calanico2020optbw}.}
	If there is no treatment heterogeneity, the first two components of our EMSE criterion are the same as this expression; otherwise, our criterion uses weighted averages. Note that in our case, there is no closed-form solution for the bandwidth as the third and fourth parts depend on the bandwidth.
	\par
	Also let us compare our criterion to the criterion developed by \cite{athey2016recursive} for discovering CATE under unconfoundedness. Their criterion does not contain the first two components, the bias square and the first variance part, while they have an unbiased estimator for the group average treatment effect by assumption. (This is similar to our parametric RD case, where we assume a parametric model for each leaf.) The third and fourth terms refer to the same quantities, and in the \href{https://github.com/regulyagoston/RD_tree/blob/main/online_appendix.pdf}{online appendix} under Section D. we show that their criterion is indeed a simplified version of ours.
	\par
	\indent 
	Our final note for the estimation is a potential further simplification. The bias-corrected variance estimator and the non-corrected versions are asymptotically the same. One can utilize this fact and simplify Equation (\ref{eq:estEMSE}) further, as Equation (\ref{eq:EMSE_exp_bc_var}) reduces to $\frac{1}{N^{te}h}\sum_{j=1}^{\nP} \hat{\mathcal{V}}_j$.\footnote{We derive a simplified formula in the \href{https://github.com/regulyagoston/RD_tree/blob/main/online_appendix.pdf}{online appendix} Section B.2.} However in our experience with finite samples, the algorithm works better if we distinguish these two measures. 
	
	\subsection{Fuzzy criterion}
	With fuzzy designs, the objective is the same; thus, the EMSE criterion is the same as in Equation (\ref{eq:sEMSE}), but its estimator differs in some aspects. In the \href{https://github.com/regulyagoston/RD_tree/blob/main/online_appendix.pdf}{online appendix} Section C, we derive the estimator, which results in the following expression,
	\begin{equation*}
		\begin{aligned}
			\widehat{EMSE}^{FRD}_\tau(\ste,\sest,\Pi,h) 
			=&  h^{2q+2} \sum_{j=1}^{\nP} \left(\hat{\mathcal{B}}^{FRD}_j\right)^2 p^{est}_j + \frac{1}{N^{est}h}\sum_{j=1}^{\nP} \hat{\mathcal{V}}^{FRD}_j\\
			&-\frac{1}{N^{te}}\sum_{i\in \ste} \left(\frac{\hat\tau^{y}(Z_i;\Pi,h,\ste)}{\hat\tau^{t}(Z_i;\Pi,h,\ste)} - \hat{\Bc}^{FRD}_j h^2\right)^2 \\
			&+ \sum_{j=1}^{\nP}p^{te}_j \V\left[\tilde\tau_j^{FRD}(h,\Pi)\right]  \, .
		\end{aligned}
	\end{equation*}
	\par
	\noindent
	Similarly to the sharp EMSE criterion, this also minimizes the bias square in the leaf-by-leaf estimates and their variance. However, let us note that the estimator for the bias square term and the expected variance over all $z=Z_i$ uses linear approximation for the CLATE parameter, similarly to \cite{calonico2014robust}, thus it does not average over the ratio of the biases for the outcome and the participation equations, but handles them jointly.
	\par
	Also, with the fuzzy design, the criterion combines the jumps in the outcome, the participation equation, and the variance. This means that if there is a difference in two groups in the participation probabilities at the threshold or in the outcome equation, then the EMSE criterion results in a lower value and finds this difference. Similarly, suppose the variance of $\hat\tau^{y}$, $\hat\tau^{t}$, or their covariance gets lower by a split. In that case, the EMSE criterion will be lower, thus even if there is no significant change in the treatment effect, but there is in its variance, the algorithm considers this split. 
	\par 
	\indent
	Finally, let us note that if the changes in the jump in the outcome and participation equations are of the same magnitude, resulting in the same treatment effect, then the EMSE criterion does not change. If one is interested in heterogeneity in the participation effect and the intent-to-treat effect separately, then it is possible to use a sharp design for both equations separately and then assemble the results from the two trees.
	\subsection{Refining honest tree algorithm for RD}
	To grow the EMSE optimal regression discontinuity tree, we follow the literature on classification and regression trees (CART) and honest causal regression trees (see, e.g., \citealt{breiman1984classification}, \citealt{ripley1996pattern} or \citealt{hastie2011elements} on CART algorithms and \citealt{athey2015machine,athey2016recursive} and \citealt{wager2018estimation} on honest causal tree algorithms).
	Here, we briefly summarize the main steps for an honest RD tree, and discuss it in more detail in the \href{https://github.com/regulyagoston/RD_tree/blob/main/online_appendix.pdf}{online appendix} under Section A.2.
	\par
	\medskip
	\noindent
	\textit{Steps for growing discontinuity tree:}
	\begin{enumerate}[topsep=4pt,itemsep=-1ex,partopsep=1ex,parsep=1ex]
		\item Split the sample into two independent parts $\to$ training and estimation samples.
		\item Grow a large tree on the training sample with bandwidth $h_r$.
		\item Prune this large tree to control for over-fitting. This is carried out by cross-validation, where we employ weakest link pruning.
		\item Minimize bandwidth $h_r$ such that the pruned tree has the smallest cross-validated EMSE value.
		\item Use this EMSE optimal tree with $h^*$ to estimate the CATE function on the independent estimation sample.
	\end{enumerate}
	\par
	\noindent
	Although the main steps are similar to the literature, here we discuss some of the refinements needed to estimate the CATE function in regression discontinuity designs.
	\par
	\noindent
	First, we offer two options to minimize the bandwidth. The default is a constrained optimization via interior point\footnote{\texttt{fmincon} function from MatLab.}. The lower bound for the bandwidth is defined such that the minimum number of effective observations is satisfied with one treatment effect. The upper bound is ten times the homogeneous optimal bandwidth (calculated with homogeneous ATE) or such that all weights are close to one. The starting point is this homogeneous optimal bandwidth for the whole population. Note that homogeneous optimal bandwidth can be MSE, or coverage error (CE) optimal bandwidth as in \cite{calanico2020optbw}. The second option is a simple grid search, where the user can set the values of the candidate bandwidths or the number of candidate bandwidths to create between minimum and maximum bandwidths, which are created similarly to those for the constrained optimization. We use the same bandwidth when we calculate the $q+1$'th order polynomials to estimate the bias and variance for the bias-correction.\footnote{This gives better performance in our simulations than when we estimate both bandwidths.}
	\par
	Another important detail is using the effective number of observations with positive weights instead of the minimum number of all observations (with zero weights included) during the estimation. This effective number of observations serves as a stopping criterion more likely to be binding. For example, suppose one has a total of 200 observations in a particular leaf, and the required effective number of observations is 50. In that case, the effective number of observations may be only 10 with a small bandwidth value. This implies that the algorithm does not consider such a candidate leaf despite having 200 observations within that leaf.
	Also, let us mention similar changes related to `bucketing'. The modified bucketing ensures that each candidate split value for a given feature has enough \textit{effective} treated and non-treated unit changes from one leaf to another. We require a splitting value to be valid compared to the last valid splitting value if it has at least four or more \textit{effective} observation changes for both treated and control groups from one subsample to the other. Using only \textit{effective} observations for both the treated and control groups allows smoothing the splitting criterion in the RD setup and considerably speeds up the computation.
	\par
	Next, let us mention that the sample size changes with each fold when we carry out cross-validation to get the optimal pruning parameter for $h_r$. We modify our bandwidth estimate for each fold by utilizing the results of \cite{calanico2020optbw}: the optimal bandwidth is proportional to the sample size.\footnote{The implemented correction depends on the initially set bandwidth type (MSE or CE). For example, when one uses CE, the adjusted bandwidth for cv-sample is $h_{r,val} = h_r \left(\frac{N_{tr,val}}{N^{tr}}\right)^{-1/(3+p)}$. This is typically not a large difference for the training set, but considerable for the evaluation set. Note that \cite{calanico2020optbw} derives the closed form solution, which is not feasible here, although it provides a good rule of thumb here. The scaling in the exponent should be between $-1/(3+q)$ and $-1/(5+2q)$ to mimic the generic bandwidth choice.}
	\par
	Finally, there is a discussion on selecting the order of polynomials used during the estimation (see, e.g., \citealt{imbens2019poly} or \citealt{pei2022local}). \footnote{The main recommendation of \cite{imbens2019poly} is to use low-order (local) polynomials to avoid noisy estimates. \cite{pei2022local} propose a measure incorporating the most frequently used nonparametric tools to select the polynomial order.}
	This paper offers a natural approach to selecting the order of polynomials: use the cross-validation procedure jointly with bandwidth and the complexity parameter to select the order of the local polynomial estimator. As the estimated EMSE value is an unbiased estimator, it will also lead to the EMSE optimal order of polynomial selection.
	\subsection{Regression discontinuity forest}
	Under the identifying assumptions from Section \ref{sec:rf-identification}, one can estimate a discontinuity forest that represents $b=1,\dots,B$ trees. To do so, one may alter the regression tree algorithm in a couple of dimensions. First, instead of create two sample, $\ste$ and $\sest$, one need $B$ bootstrapped sample with repetition, $\mathcal{S}_b$. We will use the selected in-bag sample $\mathcal{S}_b$ as $\ste_b$, and remaining observations, the out-of-bag sample $\left(\mathcal{S} \setminus \ste\right)$ as the $\sest_b$ and estimate the $b$'th tree, $\Pi_b$. Stopping criterion for the tree, along with variable selection at each node, are similar to the literature (see, e.g., \citealt{breiman2001random}, \citealt{wager2018estimation}), we grow shallower trees and randomly select a subset of the original variables, while not employing any type of pruning. We carry out cross-validation for the bandwidth and/or order of the polynomial using the out-of-bag obervations. The resulting estimator is the weighted sum estimators of the individual trees, $\hat \tau^{DF}(z;\cdot) = \sum_{b=1}^{B} w_b(z) \hat\tau_b(z;\cdot)$, with $w_b(z)$ is the weight assigned to tree $b$ for input $z$.
	To calculate the variance of the estimator we follow \cite{wager2018estimation} and use infinitesimal jackknife method.
	
	\section{Monte-Carlo simulations}
	\label{sec:simul}
\begin{figure}[H]
		\centering
		\begin{subfigure}{0.32\textwidth}
			\includegraphics[width=\linewidth]{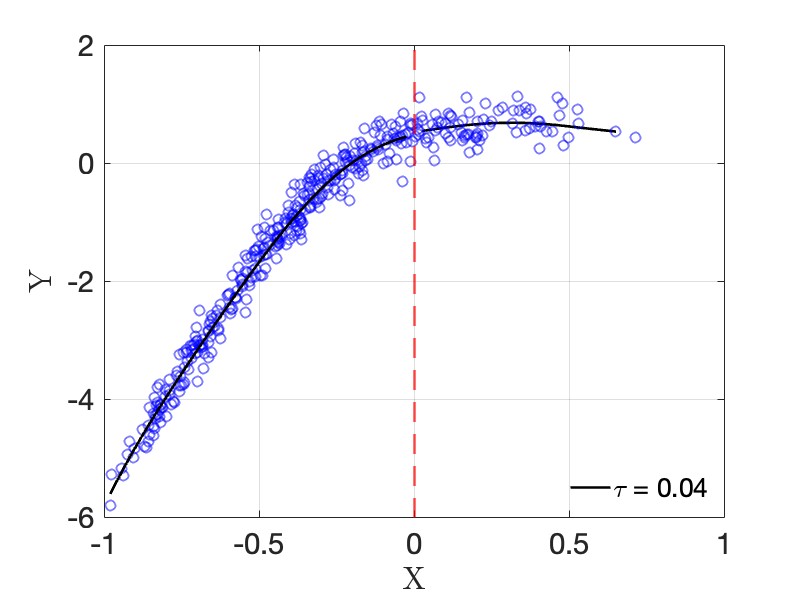}
			\caption*{\centering \footnotesize{DGP-1: One treatment effect}}
		\end{subfigure}
		\begin{subfigure}{0.32\textwidth}
			\includegraphics[width=\linewidth]{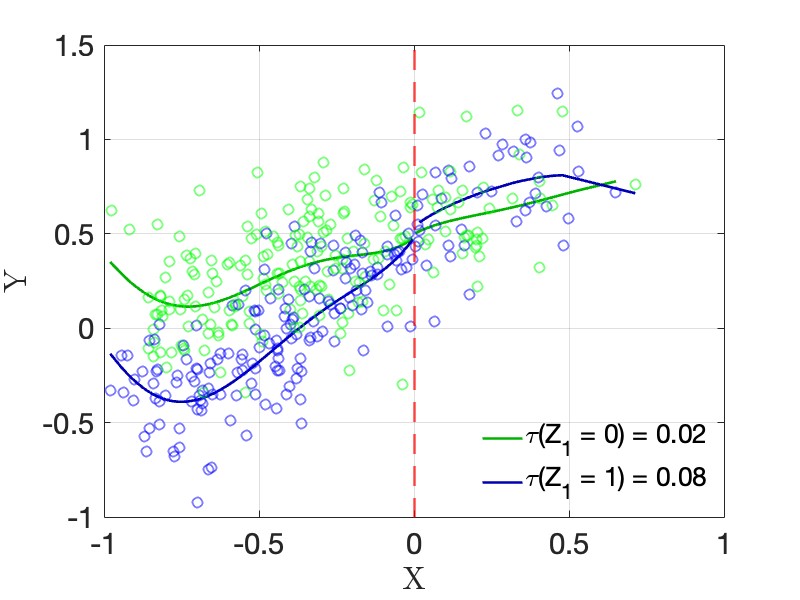}
			\caption*{\centering \footnotesize{DGP-2: Two treatment effects}}
		\end{subfigure}
		\hfill
		\begin{subfigure}{0.32\textwidth}
			\includegraphics[width=\linewidth]{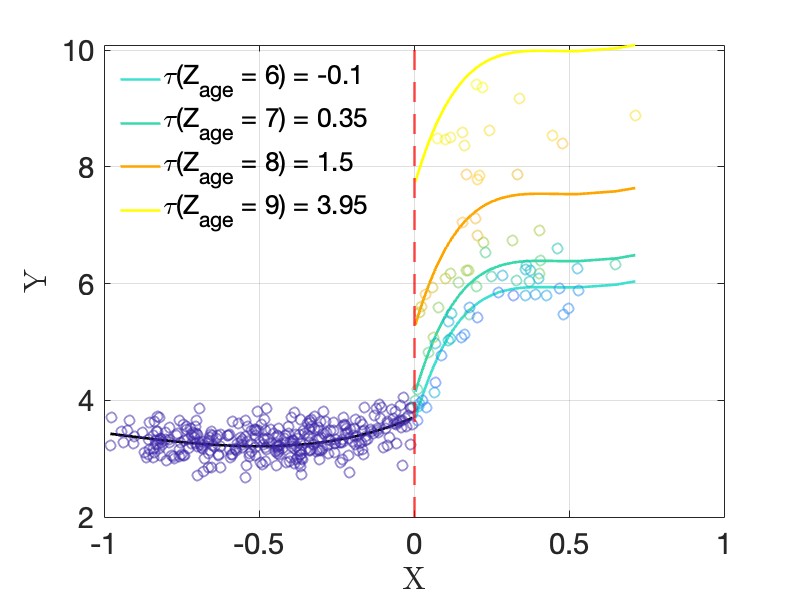}
			\caption*{\centering \footnotesize{DGP-3: Continuous CATE}}
		\end{subfigure}
		\medskip
		\caption{Monte Carlo simulation designs}
		\label{fig:mc_designs}
	\end{figure}

	To analyze the finite sample properties of the proposed methods, we follow simulation designs from \cite{calonico2014robust}, and adjust the treatment effects such that they mimic different cases: a no-heterogeneity, two distinct treatment effects and a continuous CATE.
	\begin{changemargin}{0.3in}{0in} 
		\begin{outline}
			\1 [DGP-1: ] Use ``Model 3'' from \cite{calonico2014robust} that has extra curvature in the CEF. In this design there is only one treatment effect and we added 50 dummy variables. 
			\2 $\tau = 0.04$, number of features: 50
			\1 [DGP-2: ] Imitates \cite{lee2008randomized} vote-shares application as well, but now we assume two treatment effects with different conditional expectation functions. The political party dummy $(Z_1)$ is relevant and affects both the treatment and the functional forms. We also add ``state'' variables which represents the 50 US states, which are irrelevant.
			\2 $\tau(Z_1=1)=0.02, \, \tau(Z_1=0)=0.08$, number of features: 50
			\1 [DGP-3: ] Follows \cite{ludwig2007does}, who studied the effect of Head Start funding to identify the program's effects on health and schooling. We assume a continuous treatment effect based on the age of participants ($Z_1$), while adding (irrelevant) dummies representing different continents. The CATE function is given by third order polynomial:
			\2 $\tau(Z_1)= -0.45 + 0.5Z_1 - 0.25 Z_1^2 + 0.1Z_1^3$, number of features: 7    
		\end{outline}
	\end{changemargin}
	Figure \ref{fig:mc_designs} shows the different sharp RD designs. For more detailed discussion, refer to \href{https://github.com/regulyagoston/RD_tree/blob/main/online_appendix.pdf}{online appendix}, Section E. 
	\par
	\noindent
	During the simulations, we use four different sample sizes: $N=1,000;\,5,000$, $10,000$, and $50,000$ to investigate the effect of the sample size on the algorithms. As the discontinuity tree splits the initial sample, we use half of the observations for the training and the other half for the estimation sample. We use $MC=1,000$ Monte Carlo repetitions and the variation comes from a normally distributed disturbance term, $\epsilon_i \sim \mathcal{N}(0,0.05)$, whereas the features are uncorrelated. We use the discontinuity tree algorithm along with the discontinuity forest.
	\par
	\indent
	We are using three different measures to evaluate the algorithms' performance. The first measure investigates whether the proposed estimable EMSE criterion is a good proxy to minimize the ideal infeasible criterion (Equation \ref{eq:infMSE}). We calculate this infeasible criterion for transparent comparison on a third independent evaluation sample, containing $N^{eval}=10,000$ observations. The criterion is calculated on this evaluation sample, and the CATE estimator comes from the tree or forest, which is grown on the training sample and estimated on the estimation sample. The Monte Carlo average of this estimate is reported as ``\textit{inf. MSE}''. The second measure is the average bias, of the estimated CATE measure, which shows if there is any systematic differences, Bias$:= MC^{-1}\sum_{mc=1}^{MC}\left(N^{eval}\right)^{-1}\sum_{i=1}^{N^{eval}}\left[\tau(Z_i) - \tilde{\tau}(Z_i;\Pi,h,\sest)\right]$. The third measure provides information about the empirical coverage ratio. To calculate this measure for discontinuity tree we use the tree structure (step-function approximation), thus instead of $\tau(Z_i)$, we use $\tau(Z_i;\widehat{\Pi})$ to investigate the hit ratio with 95\% confidence intervals.\footnote{This means, e.g. for DGP-2, when the algorithm only finds one treatment, for the coverage ratio we use true treatment effect that is the expected value of the combined two treatment effects. Similarly, when we have a continuous treatment effect in DGP-3, with a tree structure, we calculate the true treatment effect for each leaf and contrast this value with the estimated version. See more details in the \href{https://github.com/regulyagoston/RD_tree/blob/main/online_appendix.pdf}{online appendix} under Section E.} Finally, for the tree algorithm, we report the average number of leaves on the discovered tree $(\# \hat\Pi)$. 
	\begin{table}[H]
		\centering
		\caption{Monte Carlo averages for performance measures}
		\resizebox{\textwidth}{!}{
			\begin{tabular}{ll|cccc|ccc}
\toprule
\toprule
& & \multicolumn{4}{c|}{Tree} & \multicolumn{3}{c}{Forest}    \\
Design & Sample Size & infMSE & Bias & Coverage & $\#\Pi$ & infMSE & Bias  & Coverage  \\
\hline
\multirow{3}{*}{DGP-1} 
& $ N = 1,000$ & 0.0043 & -0.0018 & 0.8740 & 1.00 & 0.0012 & -0.0011 & 0.9370 \\ 
& $ N = 5,000$ & 0.0005 & -0.0014 & 0.9190 & 1.13 & 0.0003 & -0.0001 & 0.8745 \\ 
& $ N = 10,000$ & 0.0004 & -0.0023 & 0.9290 & 1.15 & 0.0001 & 0.0000 & 0.9155 \\ 
& $ N = 50,000$ & 0.0002 & -0.0019 & 0.9345 & 1.11 & 0.0000 & -0.0001 & 0.9458 \\  \hline
\multirow{3}{*}{DGP-2} 
& $ N = 1,000$ & 0.0050 & 0.0098 & 0.8890 & 1.00 & 0.0019 & 0.0111 & 0.8069 \\ 
& $ N = 5,000$ & 0.0009 & 0.0037 & 0.9105 & 1.41 & 0.0007 & 0.0076 & 0.7534 \\
& $ N = 10,000$ & 0.0009 & -0.0046 & 0.9035 & 1.68 & 0.0005 & 0.0029 & 0.8555 \\ 
& $ N = 50,000$ & 0.0003 & -0.0017 & 0.9490 & 1.91 & 0.0005 & -0.0013 & 0.9352 \\  \hline
\multirow{3}{*}{DGP-3} 
& $ N = 1,000$ & 1.3758 & 0.0060 & 1.0000 & 1.00 & 1.3707 & -0.0173 & 0.8537 \\ 
& $ N = 5,000$ & 0.5061 & -0.1377 & 0.9683 & 2.08 & 0.4205 & -0.1176 & 0.5369 \\ 
& $ N = 10,000$ & 0.1044 & -0.0372 & 0.8734 & 6.42 & 0.0660 & -0.0138 & 0.8848 \\ 
& $ N = 50,000$ & 0.0904 & 0.0128 & 0.9328 & 9.35  & 0.0430 & -0.0119 &  0.9330 \\  \hline
\bottomrule
\bottomrule
\end{tabular}
		}
		\begin{minipage}{.98\textwidth}
			\vspace{2pt}
			\scriptsize
			\emph{ Number of true leaves: $\nP_{DGP-1}=1$, $\nP_{DGP-2}=2$,  $\nP_{DGP-3}=Inf$. 
				Algorithm setup: using the smallest cross-validation value to select the pruning parameter $\gamma^*_{h_r}$. For bandwidth selection $(h^*)$ we use grid-search around the bandwidth for the homogeneous treatment effect with coverage error optimal bandwidth with 10 candidates. We use local linear model $(q=1)$ for each leaf and $q+1$ for bias correction. Bandwidth for higher order polynomial is selected by $\rho=1$. Leaf-by-leaf variance estimators are using the HCE-1 formula.}
		\end{minipage}
		\label{tab:mc_alg}
	\end{table}
	\noindent
	Table \ref{tab:mc_alg} shows that the algorithm works better as the sample size grows. The infeasible MSE is decreasing in $N$ for each setup. This supports the theoretical claim that the estimable EMSE criterion is a proper proxy for the infeasible MSE; thus, the resulting tree and bandwidth are MSE optimal in this sense. 
	The bias is relatively small compared to the treatment effect, decreases with the sample size, and alternates its sign. The empirical coverage ratios check for $N=50,000$, but are somewhat smaller than the theoretical 95\% with smaller sample sizes.
	For the discontinuity tree, the number of leaves $(\#\Pi)$ mimics the true number of leaves: for DGP-1, it is close to 1, for DGP-2, it increases towards 2, and for DGP-3, it increases in $N$.
	Also, let us note that the discontinuity forest produces smaller infeasible MSE values and bias, which is consistent with the literature.
	To summarize the results, the algorithm is relatively conservative in discovering different treatment effects a and data-intensive method.
	We have run simulations for fuzzy design as well. The main conclusions are similar, while the fuzzy design is even more data-intensive. See more details in the \href{https://github.com/regulyagoston/RD_tree/blob/main/online_appendix.pdf}{online appendix} under Section E.1.
	\section{Heterogeneous effect of going to a better school}
	\label{sec:emp_app}
	To show how the algorithm works in practice, we replicate and augment the heterogeneity analysis of \cite{popeleches2013} on the effect of attending a better school.
	\par
	\indent
	In Romania, a typical elementary school student takes a nationwide test in the last year of school (8th grade) and applies to a list of high schools and tracks. The admission decision is entirely dependent on the student's transition score, an average of the student's performance on the nationwide test, grade point average, and order of preference for schools.\footnote{Grades on the nationwide test are from 1-10, where 5 is the passing score on each test. Grade point average is an average of the past years' course grades for different disciplines. Order of preference for schools is a list submitted by the student before the nationwide test, showing their preferences for the schools that they apply.} A student with a transition score above a school's cutoff is admitted to the most selective school for which he or she qualifies. \cite{popeleches2013} use a large administrative dataset (more than 1.8 million observations) and a survey dataset (more than 10k observations) from Romania to study the impact of attending a more selective high school between 2003 and 2007. Based on the administrative dataset, they find that attending a better school significantly improves a student's performance on the Baccalaureate exam,\footnote{Marks in BA Exam vary from 1-10, where there are multiple disciplines, wherein each, one needs to score above five and achieve a combined score of more than 6 to pass the BA Exam.} but does not affect the exam take-up rate. 
	
	First, we revisit \cite{popeleches2013} heterogeneity analysis on the intent-to-treat effects using peer quality (level of school average transition score) and the number of schools in town as the sources of heterogeneity, using the administrative data between 2003 and 2005, while the outcome variable is the peer quality effect.
	\cite{popeleches2013} investigate treatment effect heterogeneity in two distinct dimensions: they separate students from the top and bottom terciles based on the school-level average transition score. The second dimension is the number of schools in town, and they create groups defined by having i) four or more schools in towns, ii) three schools, or iii) two schools only. \cite{popeleches2013} find significant treatment effects in all five groups. 
	\par
\begin{figure}[H]
		\centering
		\hfill
		\begin{subfigure}{0.49\textwidth}
			\includegraphics[width=\linewidth]{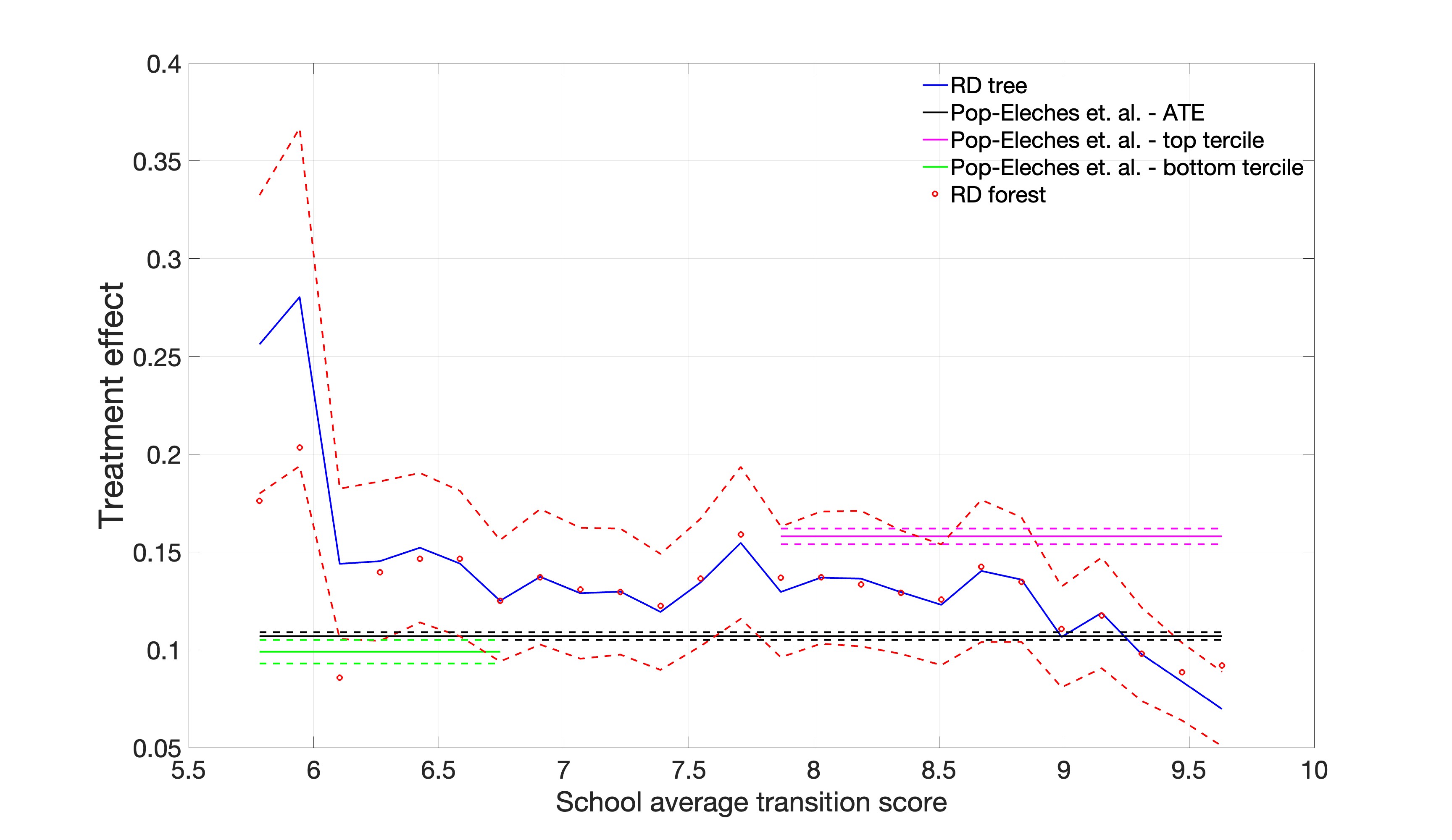}
			\caption{Level of school avg. transition score}
			\label{fig:het_agus}
		\end{subfigure}
		\begin{subfigure}{0.49\textwidth}
			\includegraphics[width=\linewidth]{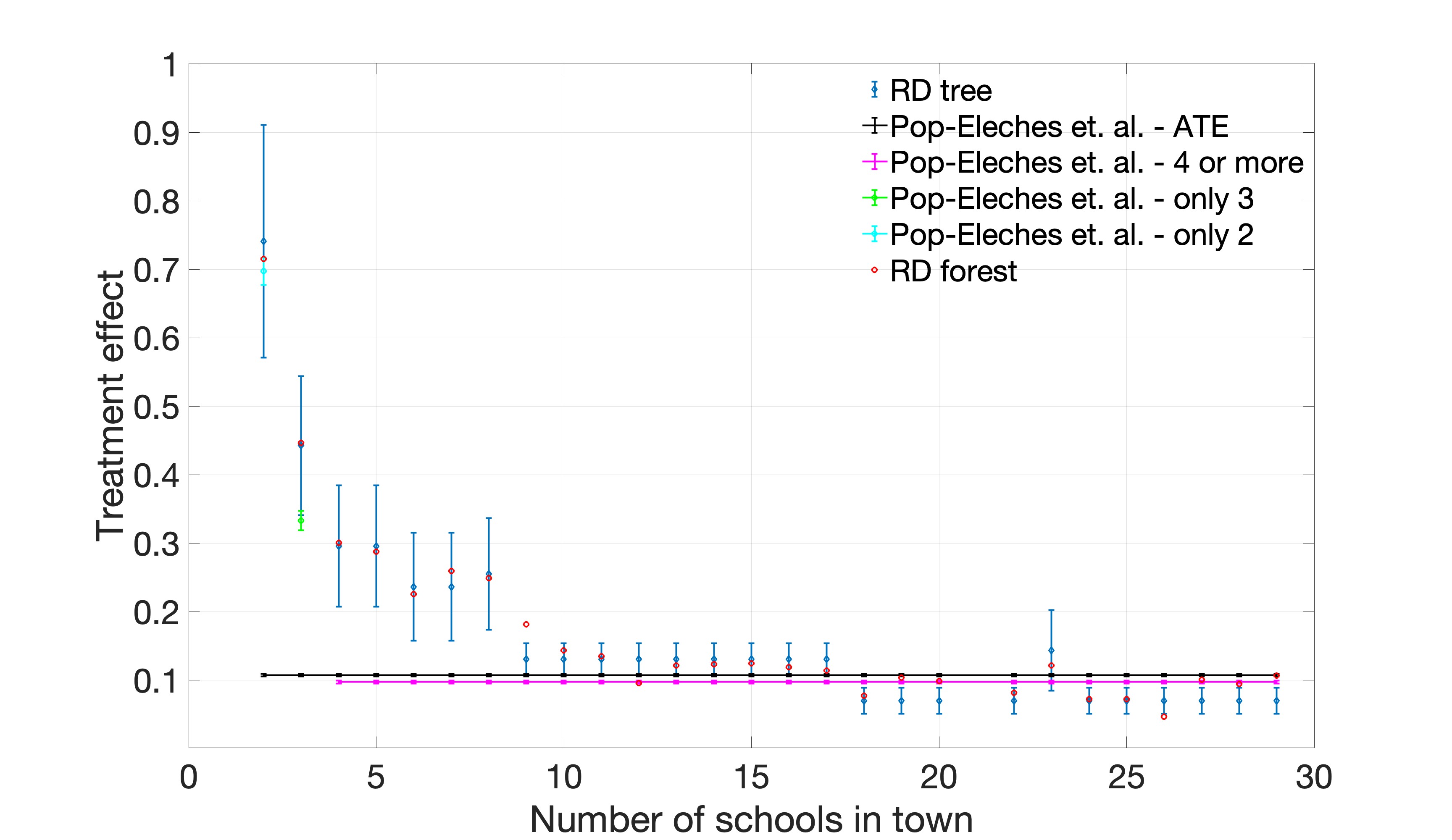}
			\caption{Number of schools}
			\label{fig:het_nschool}
		\end{subfigure}
		\caption{CATE for peer quality, intent-to-treat effects, using school fixed effects, standard errors are clustered at student level}
		\label{fig:school_hetero1}
	\end{figure}

	Instead of using these pre-specified (ad-hoc) groups, we use the algorithm to identify the relevant groups and split values.\footnote {We treat number of school variable as distinct dummy variables.} The regression discontinuity tree algorithm finds a much more detailed tree, containing 15 leaves, indicating a continuous CATE function. Instead of showing a large tree, Figure \ref{fig:school_hetero1} shows the marginalized treatment effects along the two variables.\footnote{We have calculated the treatment effect for each observation and then averaged them over the non-plotted variable. In the case of the number of schools, we take students from the same number of schools in town and average them along with the level of school average transition score.} Figure \ref{fig:het_agus}) shows the treatment effects conditional on the level of school average transition score.\footnote{We used 25 equal-sized bins to group school average values.} The blue line represents the CATE function found by the tree, and the red circles are the estimates from the forest. The black line shows the overall average treatment effect, while the green and pink lines show the treatment effects reported by \cite{popeleches2013} for the bottom and top terciles. Figure \ref{fig:het_nschool}) shows the heterogeneity in the treatment effects along with the number of schools. Similar to the previous plot, the different colored error bars show the treatment effects for the different models.
	\noindent
	It is interesting to compare the algorithm's result to \cite{popeleches2013} results (black, green and pink lines). First, the optimal unified bandwidth is 0.1481, which is different from the parametric RD that \cite{popeleches2013} uses, and result in higher overall ATE that is 0.1333 (0.0069) instead of 0.107 (0.001).\footnote{Note that these estimates are based on the estimation sample. Also, when the paper of \cite{popeleches2013} was accepted, nonparametric RD was not the standard method. They used a rule of thumb and selected all observations within the -1 to 1 regions from the cutoff. If one uses nonparametric RD on the whole sample, the resulting MSE optimal bandwidth is 0.207 and the overall ATE is 0.127 (0.003) that is higher than the reported line in Figure \ref{fig:het_agus}. See more in the \href{https://github.com/regulyagoston/RD_tree/blob/main/online_appendix.pdf}{online appendix} under Section F.} Figure \ref{fig:het_agus}) shows that the `bottom tercile' (green line) effect is below the effect what we find\footnote{When using local linear regression on this sub-sample, we found that the MSE optimal bandwidth is 0.156 and the resulting ATE is 0.117 (0.007), that is closer to our estimate.}, and the effect is the largest for the lowest school average transition scores. It is also interesting that, above 9.5, we found the lowest effects. With the discontinuity forest the results are similar. We have found lower estimates than the bottom tercile just above the value of 6 with the forest, however this is not significantly different from our estimate.
	Conditioning the treatment effects on the number of schools results in the same conclusion for two and three schools, 
	but as Figure \ref{fig:het_nschool}) shows the treatment effect suggested by the algorithm is still higher than the average for towns with 4-9 schools, and it is significantly lower for towns with 18-30 schools.
	\par
	To explore treatment effect heterogeneity and show how the algorithm performs when there are many covariates with potential non-linearities, we use the survey dataset from 2005-2007. This sample contains 11,838 observations, with a rich variety of socio-economic factors (e.g., gender, ethnicity, education, accessibility of internet or phone), school characteristics (e.g., novice teacher among teachers, highly certified teachers in schools), and study behavior-specific questions (e.g., parents pay for tutoring, parents help students, the student does homework every day, peer ranking, teacher characteristics). In the survey, there are only 135 schools located in 59 towns with 2 to 4 schools, and a questionnaire was administered between 2005 and 2007. Overall, we use 32 different features to search for heterogeneity.
	\begin{figure}[H]
		\centering
		\resizebox{\textwidth}{!}{
			\scriptsize
			\begin{tikzpicture}[align=center,
				node distance=4cm and 7cm, >=stealth]
				\Tree 
				[.{All\\0.4907 (0.0797)}
				[ .{Phone access \\ 0.6939 (0.1317)}
				[ .{ Negative interactions \\ 0.4145 (0.2360) } 
				]
				[ .{ No negative interactions \\ 0.8114 (0.1483) }
				[ .{ Age of head of household $\leq$ 48 \\ 0.6830 (0.1673) }
				]
				[ .{ Age of head of household $>$ 48 \\ 1.3806 (0.3205) }
				]
				]
				]
				[ .{No phone access \\ 0.3909 (0.0968)}
				[ .{School avg. score $\leq$ 8.10 \\ 0.0915 (0.1023)}
				]
				[ .{School avg. score $>$ 8.10 \\ 0.5907 (0.0919)}
				]
				]
				]
			\end{tikzpicture}
		}
		\caption{Conditional treatment effects for peer quality, intent-to-treat effects, using school fixed effects, standard errors are clustered at student level, $h^*=0.3973$}
		\label{fig:pop_survey}
	\end{figure}
	\noindent
	Figure \ref{fig:pop_survey} shows four relevant features with three statistically different treatment effects. First, we found no positive peer quality treatment for the subpopulation that does not have phone access, and the school average transition score is lower than the mean (8.10). This group contains 24\% of the sample. The second important subgroup that the algorithm finds is the group where there is phone access, no negative interactions with the peers, and the age of the head of household is larger than 48 years. The effect is 1.38, which is considerably higher than the average effect. This subgroup (13\% of the sample) may represent some omitted variables, such as having older siblings\footnote{The argument is that, in households where the parents are older (48+), there are probably more than one child and the observed student has older siblings, as the parents are older.} which can account for the higher positive effects. Finally, the remaining subgroups have peer quality effects between 0.41 and 0.68.
	\par
	Finally, let us relate these results to \cite{hsu2019testing}. They search for heterogeneity using peer quality as the potential source in the administrative sample and find strong evidence for the exam-taking rate (under 1\% p-values) and weak evidence for BA grade (around 10\% p-values) among schools. Although they restricted their sample to towns with two or three schools and estimated the local average treatment effect, the conclusion is the same: the values of school-level average test scores impact the treatment effect level. We have found similar results for the probability of exam taking, while no heterogeneity for the BA grade. 
	See more details in the \href{https://github.com/regulyagoston/RD_tree/blob/main/online_appendix.pdf}{online appendix} under Section F.

	\section{Conclusions}
	\label{sec:Conclusion}
	The paper proposes an algorithm that uncovers treatment effect heterogeneity in classical regression discontinuity (RD) designs. Heterogeneity is identified through the values of pre-treatment covariates, and the algorithm's task is to find the relevant groups. The introduced honest ``\textit{regression discontinuity tree}'' algorithm ensures a flexible functional form for the conditional treatment effect (CATE) with valid inference while handling many potential pre-treatment covariates and their interactions.
	\par 
	\noindent
	The identification and properties of the CATE function for sharp regression design are analyzed, and the paper shows how the algorithm works. An estimable EMSE criterion is derived, which uses the specifics of the nonparametric RD setup, such as the bandwidth selection and bias-variance trade-off. Furthermore, the algorithm utilizes the honest approach to get a valid inference for the parameter of interest. We derive the fuzzy RD setup and discuss random forest extensions.
	\par
	\noindent
	Monte Carlo simulation results show that the proposed algorithm and criterion work well and discover the true tree in more than 95\% of the cases. The estimated conditional treatment effects are unbiased, and the standard errors provide proper estimates for 95\% confidence interval coverage when there is a large enough sample.
	\par
	\noindent
	Finally, the paper shows how one can utilize the algorithm in practice. We use \cite{popeleches2013} data on the Romanian school system and uncover heterogeneous treatment effects on the impact of going to a better school. When revisiting the heterogeneity analysis done by \cite{popeleches2013}, the algorithm shows a more detailed picture.
	\bibliography{rdd_clean.bib}
	
\end{document}